\documentclass[onecolumn,showpacs,preprintnumbers,amsmath,amssymb]{revtex4}

\usepackage{graphicx}% Include figure files
\usepackage{dcolumn}% Align table columns on decimal point
\usepackage{bm}% bold math

\begin{document}

\title{Analytical Studies on a Modified Nagel-Schreckenberg Model with the Fukui-Ishibashi Acceleration Rule}
\author{Chuan-Ji Fu}
\author{Chuan-Yang Yin}
\author{Tao Zhou}
\author{Bo Hu}
\author{Bing-Hong Wang}
\email{bhwang@ustc.edu.cn, Fax:+86-551-3603574.}
\author{Kun Gao}
\affiliation{%
Nonlinear Science Center and Department of Modern Physics,
University of Science and Technology of China, Hefei, 230026, PR
China
}%

\date{\today}

\begin{abstract}
We propose and study a one-dimensional traffic flow cellular
automaton model of high-speed vehicles with the
Fukui-Ishibashi-type (FI) acceleration rule for all cars, and the
Nagel-Schreckenberg-type (NS) stochastic delay mechanism. By using
the car-oriented mean field theory, we obtain analytically the
fundamental diagrams of the average speed and vehicle flux
depending on the vehicle density and stochastic delay probability.
Our theoretical results, which may contribute to the exact
analytical theory of the NS model, are in excellent agreement with
numerical simulations.

\end{abstract}

\pacs{64.60.Ak, 89.40.+k. 02.60.Cb}

\maketitle
\section{Introduction}
Traffic flow that displays various complex behaviors is a kind of
many-body systems of strongly interacting vehicles. One of the
approaches to microscopic traffic processes is based on cellular
automaton (CA) \cite{CA}, which treats the motions of cars as
hopping processes on one-dimensional lattices. In the past few
decades, CA models for traffic flow have attracted much interest
of a community of physicists \cite{CA,Review}. Compared with other
dynamical approaches, for instance the fluid dynamical approach,
CA models are conceptually simpler, and can be easily implemented
on computers for numerical investigations \cite{ref1}.

Two popular one-dimensional (1D) CA models are the
Nagel-Schreckenberg (NS) model \cite{NS model} and the
Fukui-Ishibashi (FI) model \cite{FI}, where periodic boundary
conditions are used to mimic the traffic flow on highway. An exact
car-oriented mean field theory (COMF) have been developed to study
the FI model, with an arbitrary limit on the maximum speed
$v_{max}$, car density $\rho$, and the delay probability $f$
\cite{Wang}. So far as we know, however, there has been no
established exact analytical theory for the NS model, due to the
complications in the time evolution of the flow caused by the
acceleration and stochastic delay rules. In order to comprehend
how these rules affect the evolution and the corresponding
asymptotic steady state, we study a 1D traffic flow CA model which
combines the NS stochastic delay rule and the FI acceleration
mechanism.

This paper is organized as follows. In Sec. II, we give the
definition of the model and the evolution equations for the
inter-car spacings. Next, we present the fundamental diagrams of
the average speed and vehicle flux depending on the vehicle
density and stochastic delay probability, compared with the
numerical simulations. In Sec. IV, we summary with a discussion of
our results in connection with the FI and NS models.

\section{The Model and Analytical solution of asymptotic average Velocity}
The modified Nagel-Schreckenberg model with the Fukui-Ishibashi
acceleration rule is a probabilistic automaton, in which space and
time are discrete and hence also the velocities. The road of
length $L$ is divided into cells of certain length, each of which
can either be empty or occupied by just one car. The state of the
$n$th car ($n=1,\ldots,N$) is characterized by the momentary
velocity $v_n(t)$; in this model, the maximum speed is fixed as
$v_{max}=M$, and the vehicle density is $\rho=N/L$. Let $C_n(t)$
represent the number of empty sites in front of the $n$th vehicle
at the time step $t$, then we have
\begin{equation}
C_n(t+1)=C_n(t)+v_{n+1}(t)-v_n(t).
\end{equation}
As a function of the inter-car spacing $C_n(t)$ and the stochastic
delay probability $f$, the velocity of the $n$th car at time step
$t$ can be written as:
\begin{equation}
v_n(t)=F_M(f,C_n(t)),
\end{equation}
where
\begin{equation}
F_M(f,C)=\left\{
    \begin{array}{ccccc}
        0, &\mbox{if $C=0$}\\
        C-1 &\mbox{with probability $f$, if $0<C<M$}\\
        C &\mbox{with probability $1-f$, if $0<C<M$}\\
        M-1 &\mbox{with probability $f$, if $C\geq M$}\\
        M &\mbox{with probability $1-f$, if $C\geq M$.}
    \end{array}
    \right.
\end{equation}
Here, as one can see, we have approximately adopted the
acceleration rule introduced by Fukui and Ishibashi. The FI model
can be considered as a NS model for ``aggressive driving", since
the rules of the FI are nearly identical to the NS model, except
that the acceleration rule has been changed from ``the vehicle
speed is at most increased by 1 at each step" to ``every car
accelerates to $v_{max}$" (which is adopted in the present model)
and the stochastic delay mechanism has been modified as ``only
cars with $v_{max}$ will delay stochastically" (which is not
adopted here). For $v_{max}$=1, this may not change anything,
however, for higher velocities it will lead to a considerable
enhancement of the flow. Although the model is less realistic than
the NS, it is still of interest due to its simplicity. The
analytical description of the present model is of course much
simpler than that of the NS, and hence it might serve as a testing
ground for new analytical methods as well as new models.

If $N_i(t)$ represents the number of inter-car spacings with
length $i$ at time $t$, then the probability of finding such a
spacing at time $t$ is $P_i(t)=N_i(t)/N$. Hereafter, $P_i(t)$ will
indicated by $P_i$ for simplicity, except if specified otherwise.
Let $Q_i$ denote the probability that an arbitrary vehicle moves
$i$ sites during a given time step, then we have:
\begin{equation}
  \begin{array}{cccc}
    Q_0=P_0+fP_1\\
    Q_i=P_i(1-f)+fP_{i+1} (1\leq i\leq M-2)\\
    Q_{M-1}=P_{M-1}(1-f)+f\sum_{k=M}^{L-N} P_k\\
    Q_M=(1-f)\sum_{k=M}^{L-N} P_k.
  \end{array}
\end{equation}
Obviously, $P_i=0$ when $i>L-N$, since the maximum inter-car
spacing must be less than $L-N$. To obtain the nonvanishing $P_i$,
we introduce $W_{i\rightarrow j}$ to denote the probability of
finding an inter-car spacing with length $i$ at a given time which
changes into length $j$ at the next time. According to the Eqs.
(1)-(4), one can write all the nonzero $W_{i\rightarrow j}$ as
follows:
\begin{widetext}
\begin{equation}
  \begin{array}{cccccccc}
    W_{0\rightarrow j}=P_0Q_j,\quad (1\leq j\leq M)\\
    W_{i\rightarrow 0}=P_i(1-f)Q_0,\quad (1\leq i\leq M-1)\\
    W_{i\rightarrow j}=P_i[fQ_{j-1}+(1-f)Q_j],\quad (1\leq i\leq M-1,1\leq j\leq M,i\neq j)\\
    W_{i\rightarrow M+1}=P_ifQ_M,\quad (1\leq i\leq M-1)\\
    W_{i\rightarrow i-M}=P_i(1-f)Q_0,\quad (i\geq M)\\
    W_{i\rightarrow i-(M-j)}=P_i[fQ_{j-1}+(1-f)Q_j],\quad (i\geq M,1\leq j\leq M-1)\\
    W_{i\rightarrow i+1}=P_ifQ_M,\quad (i\geq M)\\
    W_{i\rightarrow j}=0,\quad {\rm otherwise.}
  \end{array}
\end{equation}
\end{widetext}
When the system approaches its asymptotic steady state, all the
$P_j$ will cease to change; thus the detailed statistical
equilibrium condition for the steady state holds:
\begin{equation}
\sum_{i\neq m} W_{i\rightarrow m}=\sum_{i\neq m} W_{m\rightarrow
i},\quad \forall m.
\end{equation}

Since there are $L-N+1$ variables $P_0,P_1,\cdots P_{L-N}$ in Eqs.
(6), which satisfy the following two normalization equations:
\begin{equation}
        \sum_{k=0}^{L-N} P_k=1,\quad
        \sum_{k=1}^{L-N} kP_k=\frac{1}{\rho}-1
\end{equation}
we should truncate Eqs. (6) and obtain other $L-N-1$ equations. In
the present paper, we will choose the $L-N-1$ equations with $m$
from 0 to $L-N-2$. Thus, we will have $L-N+1$ independent
equations, from which one can readily work out the values of
$P_0,P_1,\cdots P_{L-N}$, then the average speed of the traffic in
the steady state is:
\begin{widetext}
\begin{equation}
    \begin{array}{cc}
    \langle v(t\rightarrow \infty)\rangle =\sum_{i=1}^{M}
    P_i[(i-1)f+i(1-f)]+\sum_{i=M+1}^{L-N} P_i[(M-1)f+M(1-f)]\\
    =\sum_{i=1}^{M} iP_i+\sum_{i=M+1}^{L-N} MP_i-f(1-P_0)
    \end{array}
\end{equation}
\end{widetext}
Based on the above discussions, one can easily obtain the traffic
flux of the steady state:
\begin{equation}
J(t\rightarrow \infty)=\rho\langle v(t\rightarrow \infty)\rangle.
\end{equation}

\section{numerical simulations in comparison with theoretical results}
\begin{figure}
\scalebox{0.9}[0.8]{\includegraphics{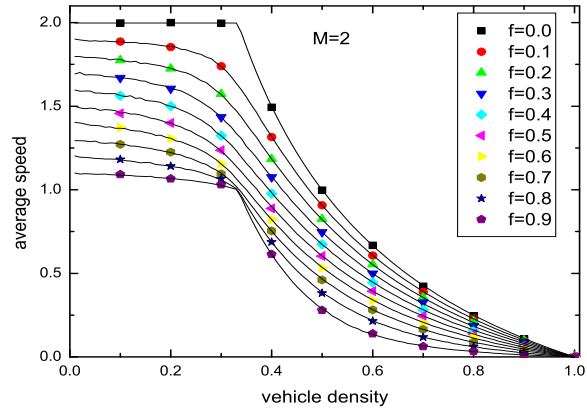}}
\caption{\label{fig:epsart} The fundamental diagram of the average
speed with the maximum speed $M=2$ and for different stochastic
delay probabilities $f$. The solid curves are numerical
simulations while the points with different symbols represent
theoretical results. The curves from the top down along the
vehicle velocity axis correspond to different values of $f$
ranging from 0 to 0.9, in steps of 0.1.}
\end{figure}

\begin{figure}
\scalebox{0.9}[0.8]{\includegraphics{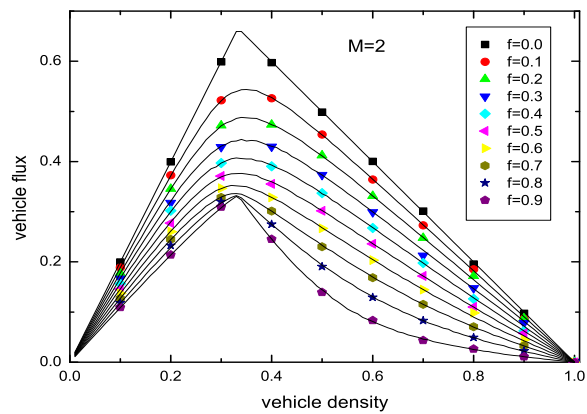}}
\caption{\label{fig:epsart} The fundamental diagram of the vehicle
flux with the maximum speed $M=2$ and for different stochastic
delay probabilities $f$. The solid curves are numerical
simulations while the points with different symbols represent
theoretical results. The curves from the top down along the
traffic flux axis correspond to different values of $f$ ranging
from 0 to 0.9, in steps of 0.1.}
\end{figure}

In order to compare with the analytical results, we performed
numerical simulations on a 1D CA with length $L=1000$ and the
maximum car velocity $M=2$. The number of vehicles $N$ is adjusted
so as to give the desired vehicle density $\rho$. The first 5000
time steps are excluded from the averaging procedure so as to
remove the transient behavior. The average values are taken over
the next 1000 time steps. Figure 1 and 2 show comparisons between
results obtained from numerical simulations and car-oriented
mean-field theory over the entire range of the vehicle density
$\rho$. The curves are the simulation  results, while the symbols
represent the theoretical results. Theoretical results are in
excellent agreement with numerical simulations.

\section{discussion}
In summary, we propose and study a one-dimensional CA model of
high-speed vehicles with the FI acceleration rule and the NS
stochastic delay mechanism. The analysis of the dynamical
evolution of our model give us a clearly physical picture of how
the acceleration and stochastic delay rules affect the evolution
and the corresponding asymptotic steady state. Although cellular
automata are designed for efficient computer simulation studies,
an analytical description is possible, though difficult all too
often. The method presented here can also be applied to other CA
models. Our study may shed some new light on developing analytical
approaches to other one-dimension traffic flow models, especially
the NS model, for which until now, no exact analytical approach
has been established. The investigation of the NS model has lead
to a better understanding of its advantages and limitations, so
that it is easier to choose the approach most suitable for a given
problem. The question, how the stationary state is approached, is
still an important open issue, which is currently under
investigation and promises to yield new insights into the physics
of the NS model.

\begin{acknowledgements}

This work has been partially supported by the State Key
Development Programme of Basic Research (973 Project) of China,
the National Natural Science Foundation of China under Grant
No.70271070 and the Specialized Research Fund for the Doctoral
Program of Higher Education (SRFDP No.20020358009)
\end{acknowledgements}


\begin{thebibliography}{Review}
\bibitem{CA} S. Wolfram, {\it Theory and Application of Cellular
Automata} (Singapore: World-Scientific, 1986); M. E. Epstein, R.
Axtell, {\it Growing artificial societies} (The MIT Press, 1996).

\bibitem{Review} D. Chowdhury, L. Santen, and A. Schadschneider,
Phys. Rep. {\bf 329}, 199 (2000); D. Helbing, Rev. Mod. Phys. {\bf
73}, 1067 (2001).

\bibitem{ref1} B.S. Kerner and P. Konhauser, Phys. Rev. E {\bf
48}, R2335 (1993); D. Helbing, Phys. Rev. E {\bf55}, 3735 (1997).

\bibitem{NS model} K. Nagel and M. Schreckenberg, J. Phys. I {\bf
2}, 2221 (1992).

\bibitem{FI} Y. Ishibashi and M. Fukui, J. Phys. Soc. Jpn. {\bf
63}, 2882 (1994); B. -H. Wang, L. Wang, and P. -M. Hui, J. Phys.
Soc. Jpn. {\bf 66}, 3683 (1997).

\bibitem{Wang} B. -H. Wang, L. Wang, P. -M. Hui, and B. -B. Hu,
Phys. Rev. E {\bf 58}, 2876 (1998); B. -H. Wang, L. Wang, P. -M.
Hui, and B. -B. Hu, Physica B {\bf 279}, 237 (2000); L. Wang, B.
-H. Wang, and B. -B. Hu, Phys. Rev. E {\bf 63}, 056117 (2001).

\end{thebibliography}
\end{document}